\documentstyle[12pt]{article}

\global\arraycolsep=1pt
\newcommand{\resection}[1]{\setcounter{equation}{0}\section{#1}}

\newcommand{\bel}[1]{\begin{equation}\label{#1}}
\newcommand{\bal}[1]{\begin{eqnarray}\label{#1}}
\newcommand{\be}{\begin{equation}}
\newcommand{\ee}{\end{equation}}
\newcommand{\ba}{\begin{eqnarray}}
\newcommand{\ea}{\end{eqnarray}}
\newcommand{\nn}{\nonumber \\}
\newcommand{\qq}{\qquad}

\renewcommand{\thefootnote}{\fnsymbol{footnote}}

\newcommand{\pz}{\partial}
\newcommand{\pbz}{\bar{\partial}}

\begin{document}

%
%
\begin{titlepage}

\begin{flushright}
       \normalsize
       June, 2002  \\
  OCU-PHYS 187\\
  KEK-TH-829 \\
  hep-th/0206123  \\

\end{flushright}

\begin{center}
{\large\bf  Normalization of Off-shell Boundary State, $g$-function \\
 and Zeta Function Regularization }
\footnote{This work is supported in part
 by the Grant-in-Aid  for Scientific Research
(14540264) from the Ministry of Education,
Science and Culture, Japan.}
 \\
\end{center}
\vfill
\begin{center}
{
{H. Itoyama${}^1$}\footnote{e-mail address:
                        {\tt itoyama@sci.osaka-cu.ac.jp}} and
{T. Oota${}^2$}\footnote{e-mail address:
                      {\tt toota@post.kek.jp}}
        }\\
\end{center}

\vfill
\begin{center}
     ${}^1$\it  Department of Mathematics and Physics,
        Graduate School of Science,\\
        Osaka City University,\\
        3-3-138, Sugimoto, Sumiyoshi-ku, Osaka 558-8585, Japan  \\

     ${}^2$\it Institute of Particle and Nuclear Studies,\\
       High Energy Research Organization (KEK),
        Tsukuba, Ibaraki 305-0801, Japan
~\\
\end{center}

\vfill


\begin{abstract}
  We consider the model in two dimensions with boundary quadratic
 deformation (BQD), which has been discussed in tachyon condensation.
The partition function of this model (BQD) on a cylinder is
 determined, using the method of zeta function regularization.
 We show that, for closed channel partition function,   
a subtraction procedure must be introduced
in order to reproduce the correct results at conformal points.
The boundary entropy ($g$-function) is determined from the
partition function  and the off-shell
boundary state.
We propose and consider a supersymmetric generalization of
BQD model, which includes a boundary fermion mass term,
and check the validity of the subtraction procedure.
\end{abstract}

\vfill

\setcounter{footnote}{0}
\renewcommand{\thefootnote}{\arabic{footnote}}

\end{titlepage}


\resection{Introduction}

Quantum field theories on manifolds
with boundaries have been studied actively in recent years.
They play important roles
in various areas of physics.  In particular,
since the proposal of Sen's conjecture on tachyon
condensation \cite{S},
the boundary string field theory \cite{W,Sh}
has received a considerable attention 
\cite{gerasimov-shatashvili, kutasov-marino-moore}.
With regard to this last development, the two-dimensional off-critical model with
 boundary quadratic deformation (BQD) has been considered.
It describes an off-shell renormalization group flow
from the Neumann boundary condition to the Dirichlet boundary
condition.
The closed channel partition function on a cylinder
has been considered by using the method of thermodynamic Bethe ansatz
 in the long cylinder limit \cite{FI}. 
See, for instance, \cite{Su, BK, APS} for related calculation.

In this paper, taking the BQD model with more general boundary
conditions,
we reexamine the partition function
by another method
without taking the long cylinder limit.
In order to get a finite expression for the closed channel
partition function,
we propose a subtraction procedure which follows
the zeta function regularization.
 From the expression of the partition function,
the $g$-function is determined
with the help of an off-shell boundary state \cite{FI, Lee}.

To check the validity of our subtraction procedure,
we also consider a supersymmetric generalization of the BQD model.
The BQD model can be
considered as a weak interaction limit of the
boundary sine-Gordon model \cite{GZ}.
We propose a supersymmetric generalization of BQD model
(SBQD model) of tachyon condensation
 \cite{gerasimov-shatashvili, kutasov-marino-moore}
 as the weak interaction limit of
a supersymmetric boundary sine-Gordon model.
 (For the bosonic counterpart of this statement, see
 \cite{harvey-kutasov-Martinec, Fujii-Itoyama-1} .)

Let us briefly recall the supersymmetric sine-Gordon model
on a half-line \cite{IOZ,N}.
The following action is conjectured to be integrable
\cite{N} (see, in particular the note added)
\be
S =  \frac{1}{2\pi} \int_{-\infty}^{\infty} {\rm d}\sigma^2
\int_{-\infty}^0 {\rm d}\sigma^1 {\cal L}_0
+ \frac{1}{4\pi} \int_{-\infty}^{\infty} {\rm d}\sigma^2
{\cal L}_b,
\ee
where
\ba
{\cal L}_0 &=& \frac{2}{\alpha'} \pbz X \pz X + \psi \pbz \psi
+ \tilde{\psi} \pz \tilde{\psi}
- \frac{m^2}{\lambda^2 \alpha'} \left( \cos \lambda X - 1 \right)
+ i m \tilde{\psi} \psi \cos \frac{\lambda}{2} X, \cr
{\cal L}_b &=&  a \partial_2 a
+ \omega_0^{-1} \lambda \rho^{1/2} a (\psi - i\tilde{\psi})
\cos \frac{\lambda}{4} \left( X - \chi_0 \right) \nn
& & \qq
- \frac{4}{\alpha'} \left(
\Lambda \cos \frac{\lambda}{2} \left( X - X_0 \right) - \rho
\right)
+ i \tilde{\psi}\psi \Bigr|_{\sigma^1=0}.
\ea
Here $\omega_0=e^{i \pi/4}$, $z=\sigma^1 + i \sigma^2$ and
$a=a(\sigma_2)$ is an auxiliary fermionic field.
The parameters $\rho (>0)$ and $\chi_0$ are related to $\Lambda (>0)$ and $X_0$
by the conditions:
\be
\rho \sin \frac{\lambda}{2} \chi_0
= \Lambda \sin \frac{\lambda}{2} X_0 , \qq
\rho \cos \frac{\lambda}{2} \chi_0
= \Lambda \cos \frac{\lambda}{2} X_0
- \frac{2m}{\lambda^2}.
\ee
Let us consider the massless case $m=0$. Then $\rho = \Lambda$ and
$\chi_0 = X_0$, and we find
\ba
{\cal L}_0 &=& \frac{2}{\alpha'} \pbz X \pz X + \psi \pbz \psi
+ \tilde{\psi} \pz \tilde{\psi}, \nn
{\cal L}_b &=& a \partial_2 a
+ \omega_0^{-1}
\lambda \Lambda^{1/2} a ( \psi - i \tilde{\psi} )
-   \frac{4\Lambda}{\alpha'}
\left( \cos \frac{\lambda}{2} (X- X_0) - 1 \right)
+ i  \tilde{\psi} \psi \Bigr|_{\sigma^1=0}.
\ea
If we take the $\lambda=0$ limit with $h = \lambda \Lambda^{1/2}$ fixed,
then the boundary Lagrangian becomes
\be
{\cal L}_b =
a \partial_2 a
+ \omega_0^{-1}h a ( \psi - i \tilde{\psi} )
+ \frac{h^2}{2\alpha'} (X-X_0)^2
  + i \tilde{\psi} \psi \Bigr|_{\sigma^1=0}.
\ee
The bosonic sector is the BQD
model and the fermionic sector is the Ising model
with the boundary magnetic field \cite{GZ,C,LMSS}.
  This action differs from the one in 
\cite{harvey-kutasov-Martinec, kutasov-marino-moore} by the presence of
 the boundary fermion mass term.
Inspired by this result, we propose that the action of SBQD
model on a cylinder is given by eq.(\ref{ASBQD}).
As for the Ising model on a cylinder with boundary magnetic fields,
 the $g$-functions are calculated by other methods
in the long cylinder length limit \cite{C,LMSS}.
By using our subtraction procedure,
we determine the $g$-functions. The results
 agree with those of \cite{C,LMSS}. Also, some of the results
 reproduce those of \cite{BK, APS}.

This paper is organized as follows.
The BQD model on the cylinder is considered in section 2.
The SBQD model is examined in section 3.
The section 4 is devoted to discussion.


\resection{BQD model on a cylinder}

In two dimensional Euclidean space,
the action of
the boundary quadratic deformation (BQD) model on a cylinder
is given by
\be
S[X] = \int_0^{2\pi r} {\rm d}\sigma^2 \ L,
\ee
where
\be
L = \frac{1}{4\pi \alpha'}
\left[ \int_0^l {\rm d}\sigma^1\ (\partial_a X)^2
+ v(X-X_0)^2 \Bigr|_{\sigma^1=0} +  v' (X-X_0')^2 \Bigr|_{\sigma^1=l}
\right].
\ee
We consider the cylinder of the length $l$
and of circumference $2\pi r$.

The boundary conditions at $\sigma^1=0$ and $\sigma^1=l$ are of mixed type:
\be
\partial_1 X - v \left( X - X_0 \right) \Bigr|_{\sigma^1=0} = 0, \qq
\partial_1 X + v' \left( X - X_0' \right) \Bigr|_{\sigma^1=l} = 0.
\ee
We denote the mixed type boundary condition by $B=B(u)$ where
$u= rv$.
Also, we denote the Dirichlet type boundary condition by $D=B(\infty)$
and the Neumann type boundary condition by $N=B(0)$.

We can expand
\be
X(\sigma^1,\sigma^2) = \hat{X}_0(\sigma^1) + \tilde{X}(\sigma^1, 
\sigma^2)
\ee
where the zero eigenvalue function is given by
\be
\hat{X}_0(\sigma^1) = \frac{1}{w+w'+ww'}
\left[ w w'(X_0' - X_0 ) (\sigma^1/l) + ( w + w w') X_0 + w' X_0' 
\right].
\ee
Here $w=vl$ and $w'= v'l$.
By substituting the expansion
into the Lagrangian, we have
\be
L
=\frac{1}{4\pi \alpha'} \int_0^l {\rm d}\sigma^1\
\left[\left(\partial_2 \tilde{X} \right)^2
+ \tilde{X} ( - \partial_1^2 ) \tilde{X} \right]
+ \frac{1}{4\pi \alpha' l}
\frac{w w'}{(w + w' + w w')}
\left( X_0 - X_0' \right)^2.
\ee
The oscillating modes are expressed as
\be
\tilde{X}(\sigma^1, \sigma^2) = \sum_{j=1}^{\infty} X_j(\sigma^2) 
f_j(\sigma
^1),
\ee
where
\ba
f_j(\sigma^1) &=& (-1)^j \sqrt{\frac{\pi \alpha'}{l \rho(\nu_j)}}
\left[
\frac{(\pi \nu_j - i w)}{\sqrt{(\pi \nu_j)^2 + w^2}}
e^{i\pi \nu_j \sigma^1/l}
+ \frac{(\pi \nu_j + i w)}{\sqrt{(\pi \nu_j)^2 + w^2}}
e^{-i\pi \nu_j \sigma^1/l}
\right] \nn
&=& -  \sqrt{\frac{\pi \alpha'}{l \rho(\nu_j)}}
\left[
\frac{(\pi \nu_j + i w')}{\sqrt{(\pi \nu_j)^2 + (w')^2}}
e^{i\pi \nu_j (\sigma^1-l)/l}
+ \frac{(\pi \nu_j - i w')}{\sqrt{(\pi \nu_j)^2 + (w')^2}}
e^{-i\pi \nu_j (\sigma^1-l)/l}
\right].
\ea
Here the density function is given by
\be
\rho(k) = 1 + \frac{w}{\pi^2 k^2 + w^2}
+ \frac{w'}{\pi^2 k^2 + (w')^2},
\ee
and the constants $\nu_j$ are the positive solutions of the
following relation \cite{APS}:
\bel{gbc1}
e^{2\pi i \nu_j}
\frac{(\pi \nu_j - iw)}{(\pi \nu_j + i w)}
\frac{(\pi \nu_j - iw')}{(\pi \nu_j + i w')} = 1.
\ee
Here the normalization is chosen such that
\be
\frac{1}{2\pi \alpha'} \int_0^l {\rm d}\sigma^1 \
f_j(\sigma^1) f_k(\sigma^1) = \delta_{jk},
\ee
and we have arranged the solutions in increasing order:
\be
0 \leq \nu_1 < \nu_2 < \ldots < \nu_j < \nu_{j+1} < \ldots.
\ee
Then
\bel{Lag1}
L = \sum_{j=1}^{\infty} \frac{1}{2}
\left[
(\partial_2 X_j)^2 + \left( \frac{\pi \nu_j}{l} \right)^2
X_j^2 \right]
+ \frac{1}{4\pi \alpha' l}
\frac{w w'}{(w + w' + w w')}
\left( X_0 - X_0' \right)^2.
\ee


\subsection{Some Properties of $\{ \nu_j \}$}

For later convenience,
let us rewrite eq.(\ref{gbc1}) in the following form:
\bel{gbc2}
F_{BB'}^{(-)}(\nu_j) = 0,
\ee
where
\ba
F_{BB'}^{(-)}(k)
&=& \frac{1}{2} \left[
1 - e^{2\pi i k}
\left( \frac{\pi k - i w}{\pi k + i w} \right)
\left( \frac{\pi k - i w'}{\pi k + i w'} \right)
\right] \nn
&=&\frac{i e^{\pi i k} \pi k}{(\pi k + i w)(\pi k + i w')}
\left[ (w+w')\cos(\pi k) + ( w  w' - \pi^2 k^2 ) \frac{\sin (\pi k)}{\pi 
k}
\right].
\ea
Here $B=B(u)$ and $B'=B(u')$. ($u'=rv'$.)

Let us examine the following entire function:
\be
F_-(k) :=
\left[ (w+w')\cos(\pi k) + ( w  w' - \pi^2 k^2 ) \frac{\sin (\pi k)}{\pi 
k}
\right].
\ee
It is an even function and  has zeros at $k= \pm \nu_j$.
$F_-(0)=(w+w'+ww')$.
The key observation is that it can be written as an infinite product:
\be
F_-(k) = ( w + w' + w w') \prod_{j=1}^{\infty}
\left[ 1 - \frac{k^2}{\nu_j^2} \right].
\ee
Therefore
\bal{FmBB}
F_{BB'}^{(-)}(k) &=&
\frac{i e^{\pi i k} \pi k}{(\pi k + i w)(\pi k + i w')}
\left[
( w + w' ) \cos( \pi k)
- \left\{ (\pi k)^2 - w w' \right\} \frac{\sin (\pi k)}{\pi k}
\right] \nn
&=& \frac{i e^{\pi i k} \pi k ( w + w' + w w' )}
{ (\pi k + i w )(\pi k + i w' )}
\prod_{j=1}^{\infty}
\left[ 1 - \frac{k^2}{\nu_j^2} \right].
\ea

At conformal points, $\nu_j$ can be written explicitly.

(i) For $DD$ boundary condition ($w=\infty$ and $w'=\infty$),
\be
F_{DD}^{(-)}(k)
=
\frac{1}{2} \left( 1 - e^{2\pi k} \right)
= -i e^{\pi i k} \sin (\pi k)
= -i e^{\pi i k} \pi k
\prod_{j=1}^{\infty} \left[
1 - \frac{k^2}{j^2} \right],
\ee
\be
\nu_j = j, \qq j=1, 2, \ldots.
\ee
(ii)  For $DN$ ($w=\infty$ and $w'=0$) or $ND$ ($w=0$ and $w'=\infty$),
\be
F_{DN}^{(-)}(k) = F_{ND}^{(-)}(k)
=
\frac{1}{2} \left( 1 + e^{2\pi k} \right)
= e^{\pi i k} \cos(\pi k)
= e^{i \pi k}
\prod_{j=1}^{\infty} \left[
1 - \frac{k^2}{(j-1/2)^2} \right],
\ee
\be
\nu_j= j - \frac{1}{2}, \qq j=1, 2, \ldots.
\ee
(iii)  For $NN$ ($w=0$ and $w'=0$),
\bel{FmNN}
F_{NN}^{(-)}(k)
= \frac{1}{2} \left( 1 - e^{2\pi i k} \right)
= -i e^{\pi i k} \sin (\pi k)
= -i e^{\pi i k} \pi k
\prod_{j=2}^{\infty} \left[
1 - \frac{k^2}{(j-1)^2} \right].
\ee
\be
\nu_j= j - 1, \qq j=1, 2, \ldots.
\ee
Other limits are
\ba
F_{BD}^{(-)}(k)
&=&
\frac{1}{2} \left[
1 + e^{2\pi i k} \left( \frac{\pi k - i w}{\pi k + i w} \right)
\right]
= \frac{e^{i\pi k} \pi k ( 1 + w)}{(\pi k + i w)}
\left[ \cos( \pi k) + w \frac{\sin (\pi k)}{\pi k} \right] \nn
&=& \frac{e^{i\pi k} \pi k ( 1 + w)}{(\pi k + i w)}
\prod_{j=1}^{\infty} \left[ 1 -\frac{k^2}{\nu_j^2} \right].
\ea
\ba
F_{BN}^{(-)}(k)
&=&
\frac{1}{2} \left[
1 - e^{2\pi i k} \left( \frac{\pi k - i w}{\pi k + i w} \right)
\right]
= \frac{e^{i\pi k} iw }{(\pi k + i w)}
\left[ \cos( \pi k) - \frac{\pi k}{w} \sin (\pi k) \right] \nn
&=& \frac{e^{i\pi k} i w }{(\pi k + i w)}
\prod_{j=1}^{\infty} \left[ 1 -\frac{k^2}{\nu_j^2} \right].
\ea
Comparing the expression (\ref{FmBB}) and (\ref{FmNN}),
cancellation of two poles and two zeros should occur
for $w\rightarrow 0$ and $w'\rightarrow 0$ limit:
\be
\lim_{w, w' \rightarrow 0}
\frac{(w+w' + ww')}{(\pi k + i w)(\pi k + i w')}
\left[ 1 - \frac{k^2}{\nu_1^2} \right]
= -1.
\ee
Thus for small $w$ and $w'$, the smallest solution $\nu_1$ approaches
zero as
\bel{zeronu1}
\pi \nu_1 \sim \sqrt{w + w' + w w'} \sim \sqrt{w + w'}.
\ee


\subsection{The path integral approach}

Let us evaluate the partition function
\be
Z_{BB'} = \int {[} dX {]}\  \exp\left( - S[X] \right).
\ee

We can quantize the Lagrangian (\ref{Lag1})
by the standard method of functional integrations
 (see e.g. Appendix A of \cite{P}).
 We adopt a different approach which will be convenient
for our purposes.
In this approach,
we impose the periodic boundary condition in $\sigma^2$-direction
and we further expand the oscillating modes as
\be
X_j(\sigma^2) = \sum_{m \in {\bf Z}} {\cal A}_{m,j} \ c_m(\sigma^2),
\ee
where
$c_0(\sigma^2) = (2\pi r)^{-1/2}$,
and
$c_m(\sigma^2) = (\pi r)^{-1/2} \sin(m\sigma^2/r)$,
$c_{-m}(\sigma^2) = (\pi r)^{-1/2} \cos( m\sigma^2/r)$
for $m>0$.

Note that
\be
\int_0^{2\pi r } {\rm d}\sigma^2 \ c_m(\sigma^2) c_n(\sigma^2)
= \delta_{mn}, \qq
m, n \in {\bf Z}.
\ee
The action $S$ becomes
\be
S = S_{{\rm cl}}
+ \sum_{m \in {\bf Z}} \sum_{j=1}^{\infty} \frac{1}{2}
\left[ \left( \frac{m}{r} \right)^2
+ \left( \frac{\pi \nu_j}{l} \right)^2 \right]
\left( {\cal A}_{m,j} \right)^2,
\ee
where
\be
S_{{\rm cl}} =  \frac{\tau_2}{2\pi \alpha'}
\frac{ww'}{w + w' + w w'}\left( X_0 - X_0' \right)^2, \qq
\tau_2 = \pi r/l.
\ee
Then the partition function is given by
\be
Z_{BB'}(X_0 - X_0')
= \exp\left( - S_{{\rm cl}} \right)
Z_{BB'}(0),
\ee
where
\be
Z_{BB'}(0) = {\rm Det}^{-1/2}
\left( - \partial_1^2 - \partial_2^2 \right)
= \left(
\prod_{m \in {\bf Z}} \prod_{j=1}^{\infty}
\left[ (m/r)^2 + ( \pi \nu_j/l)^2 \right] \right)^{-1/2}.
\ee

Formally, this divergent quantity can be written as
\be
Z_{BB'}(0)
= \prod_{j=1}^{\infty} \left( \frac{l}{\pi \nu_j } \right)
\cdot \prod_{m=1}^{\infty} \prod_{j=1}^{\infty}
\left[ (m/r)^2 + (\pi \nu_j/l)^2 \right]^{-1}.
\ee

We apply the zeta-function regularization naively
for each infinite product.
We regularize the partition function in two steps.
Applying the zeta-function regularization in
different order, we will obtain two equivalent expressions for the
partition function.

First, we start from the $m$-product.
We regularize
\be
\prod_{m=1}^{\infty} \left[
(m/r)^2 + (\pi \nu_j/l)^2 \right]
= \prod_{m=1}^{\infty} \left(\frac{m}{r} \right)^2
\cdot \prod_{m=1}^{\infty}
\left[ 1 + \left( \frac{\tau_2 \nu_j}{m} \right)^2 \right]
\ee
into
\be
\left( \frac{l}{\pi \nu_j} \right)
  q^{-\nu_j/2} \left( 1 - q^{\nu_j} \right),
\ee
where
$q = e^{-2\pi \tau_2}$ and $\tau_2 = \pi r/l$.
Then we get still divergent expression of the partition function:
\be
Z_{BB'}(0) = \prod_{j=1}^{\infty}
q^{\nu_j/2}\left( 1 - q^{\nu_j} \right)^{-1}.
\ee
The divergence comes from the infinite Casimir energy.

We regularize the Casimir energy
and obtain the first regularized expression:
\bel{ZBB1}
Z_{BB'}^{(1)}(0) =
q^{(1/2)c^{(-)}_{BB'}}
\prod_{j=1}^{\infty} \left( 1 - q^{\nu_j} \right)^{-1},
\ee
where
\be
c_{BB'}^{(-)}
= \int_0^{\infty} {\rm d}t \ 2t\ \rho(it)
\left[ 1 - e^{2\pi t}
\left( \frac{\pi t + w}{\pi t - w} \right)
\left( \frac{\pi t + w'}{\pi t - w'} \right)
\right]^{-1}.
\ee
See appendix for details.

Secondly, we regularize the partition function from the $j$-product.
Note that
\be
\prod_{j=1}^{\infty} \left[
(m/r)^2 + ( \pi \nu_j/l)^2 \right]
= \prod_{j=1}^{\infty}
\left( \frac{\pi \nu_j}{l} \right)^2 \cdot
\prod_{j=1}^{\infty}
\left[ 1 + \frac{(\tilde{\tau}_2 m)^2}{\nu_j^2} \right],
\ee
where $\tilde{\tau}_2 = l/\pi r = 1/\tau_2$.
Substituting $k=i\tilde{\tau}_2 m$ into (\ref{FmBB}), we have an identity
\be
\prod_{j=1}^{\infty}
\left[ 1 + \frac{(\tilde{\tau}_2 m)^2}{\nu_j^2} \right]
= \frac{(m+u)(m+u')}{2m(u + u' +\pi \tilde{\tau}_2 u u')}
\tilde{q}^{-(1/2)m}
\left[
1 - \left( \frac{m-u}{m+u} \right)
\left( \frac{m-u'}{m+u'} \right) \tilde{q}^m \right],
\ee
where $u = rv$, $u' = rv'$
and $\tilde{q} = e^{-2\pi \tilde{\tau}_2 }$.
We need to regularize the following infinite product
\be
G(u,u'):= \prod_{m=1}^{\infty}
\frac{(m+u)(m+u')}{ 2m ( u + u' + \pi \tilde{\tau}_2 u u')}
= \frac{e^{-\gamma (u+u')}}{\Gamma(u+1)\Gamma(u'+1)}
\prod_{m=1}^{\infty}
\left\{
\frac{m e^{(u+u')/m}}{ 2(u + u' + \pi \tilde{\tau}_2 u u')}
\right\}.
\ee
Here $\gamma$ is the Euler's constant.

We regularize the divergent sum as
\be
\sum_{m=1}^{\infty} \frac{u}{m} \rightarrow
\sum_{m=1}^{\infty} \left( \frac{u}{m} \right)^s
= u^s \zeta(s), \qq {\rm Re}\ s >1.
\ee
As $s$ approaches $1$, the regularized quantity behaves as
\ba
u^s \zeta(s) &=& \frac{u}{s-1} + \gamma u + u \log u + {\cal O}(s-1) \nn
&=& \frac{s u}{s -1} + \gamma u + u \log u - u + {\cal O}(s-1).
\ea
Our subtraction procedure is to replace the divergent sum
$\sum (u/m)$ by
\be
\lim_{s \rightarrow 1} \left( u^s \zeta(s) - \frac{s u}{s-1} \right)
  = \gamma u + u \log u - u.
\ee
Then the resulting expression is given by
\be
G(u, u') = \frac{\sqrt{4\pi}(u+u' + \pi \tilde{\tau}_2 u u')^{1/2}}
{\Gamma(u+1)\Gamma(u'+1)}
\left( \frac{u}{e} \right)^u\left( \frac{u'}{e} \right)^{u'}.
\ee
The subtraction part is chosen such that the 
expression has consistent zeta-function
regularization relation for large $u$, $u'$. For example,
$G(\infty, \infty) = \prod_{m=1}^{\infty} (2 \pi \tilde{\tau}_2 m)^{-1}
= (\tilde{\tau}_2 )^{1/2}$.

Then, finally we have another expression:
\bal{ZBB2}
Z_{BB'}^{(2)}(0)
&=&
\frac{1}{\sqrt{4\pi}}
\frac{\Gamma(1+u) \Gamma(1+u')}
{(u+u' + \pi \tilde{\tau}_2 u u')^{1/2}}
\left( \frac{e}{u} \right)^u
\left( \frac{e}{u'} \right)^{u'} \nn
& & \qq
\times
\tilde{q}^{-1/24}
\prod_{m=1}^{\infty}
\left( 1 - \left( \frac{m-u}{m+u} \right)
\left( \frac{m-u'}{m+u'} \right) \tilde{q}^m \right)^{-1}.
\ea

Therefore we conjecture
the equality of the two expression of $Z_{BB'}(0)$ (\ref{ZBB1})
and (\ref{ZBB2}) : $Z_{BB'}^{(1)}(0) = Z_{BB'}^{(2)}(0)$.
This is consistent with \cite{BK,APS}.

We can check our conjecture at conformal points. Indeed,
\be
Z_{DD}^{(1)}(0) = q^{-1/24}
\prod_{n=1}^{\infty} ( 1 - q^n)^{-1}, \qq
Z_{DD}^{(2)}(0) = \frac{1}{\sqrt{\tilde{\tau}_2}}
\tilde{q}^{-1/24}
\prod_{n=1}^{\infty}( 1 - \tilde{q}^n )^{-1},
\ee
\be
Z_{ND}^{(1)}(0) = q^{1/48}\prod_{n=1}^{\infty}
\left( 1 - q^{(n-1/2)} \right)^{-1}, \qq
Z_{ND}^{(2)}(0) = \frac{1}{\sqrt{2}}
\tilde{q}^{-1/24}
\prod_{n=1}^{\infty} \left( 1 + \tilde{q}^n \right)^{-1}.
\ee
For $NN$ case $\nu_1$ becomes zero, so we set
\be
\hat{Z}_{NN}^{(i)}(0):=
\lim_{u,u'\rightarrow 0} ( 1 - q^{\nu_1}) Z_{BB'}^{(i)}(0),
\qq i=1, 2.
\ee
Using eq.(\ref{zeronu1}), we have
\be
\hat{Z}_{NN}^{(1)}(0) = q^{-1/24}
\prod_{n=1}^{\infty} ( 1 - q^n)^{-1}, \qq
\hat{Z}_{NN}^{(2)}(0) = \frac{1}{\sqrt{\tilde{\tau}_2}}
\tilde{q}^{-1/24}
\prod_{n=1}^{\infty}( 1 - \tilde{q}^n )^{-1}.
\ee
Thus we see that in the limiting cases we have obtained the correct results:
$Z_{DD}^{(1)}(0) = Z_{DD}^{(2)}(0)$, $Z_{ND}^{(1)}(0) = Z_{ND}^{(2)}(0)$,
$\hat{Z}_{NN}^{(1)}(0)=\hat{Z}_{NN}^{(2)}(0)$.


\subsection{Boundary states and $g$-function}

Let us express $Z_{BB'}$ by using boundary states:
\be
\Bigl. \left( \partial_1 X - v (X-X_0) \right)
\Bigr|_{\sigma^1=0} | B(u); X_0 \rangle =0.
\ee
The mode expansion of $X$ is given by
\be
X(\sigma^1,\sigma^2) = x - \frac{i\alpha'}{r} p \sigma^1 +
i \sqrt{\frac{\alpha'}{2}}
\sum_{n \neq 0} \frac{1}{n} \left(
\alpha_n e^{-n(\sigma^1+i\sigma^2)/r}
+ \tilde{\alpha}_n e^{-n(\sigma^1-i\sigma^2)/r} \right).
\ee
and the commutation relations are
\be
{[} x, p {]} =i, \ \
{[} \alpha_m, \alpha_n {]} = m \delta_{m+n,0}, \ \
{[} \tilde{\alpha}_m, \tilde{\alpha}_n {]} = m \delta_{m+n}, \ \
{[} \alpha_m, \tilde{\alpha}_n {]} = 0.
\ee
Our normalization for the momentum and position eigenstates are
\be
\langle k | k' \rangle = \delta( k - k'), \ \
\langle X | X' \rangle = \delta( X - X'), \ \
\langle X | k \rangle = \frac{1}{\sqrt{2\pi}} e^{ikX}.
\ee
The oscillator vacuum $|0 \rangle_{\alpha}$ is normalized as
${}_{\alpha} \langle 0 | 0 \rangle_{\alpha} = 1$.
The vacuum state is denoted by
$| 0 \rangle = | k=0 \rangle \otimes | 0 \rangle_{\alpha}$.
The Hamiltonian of the system is
\be
H = \frac{1}{r} \left( L_0 + \tilde{L}_0 - \frac{1}{12} \right),
\ee
where
\be
L_0 = \frac{\alpha'}{4} p^2 + \sum_{n>0} \alpha_{-n} \alpha_n, \qq
\tilde{L}_0 = \frac{\alpha'}{4} p^2 +
\sum_{n>0} \tilde{\alpha}_{-n} \tilde{\alpha}_n.
\ee

The boundary state is given by
\ba
& & | B(u) ; X_0 \rangle \nn
&=& {\cal N}(u)
\exp \left( - \frac{u}{2\alpha'} (x-X_0)^2 \right)
\exp\left( - \sum_{n=1}^{\infty} \frac{1}{n}
\left( \frac{n-u}{n+u} \right) \tilde{\alpha}_{-n} \alpha_{-n} \right)
| 0 \rangle,
\ea
where
\be
{\cal N}(u) = ( 2 \alpha' )^{-1/4} \left( \frac{e}{u} \right)^u
\Gamma( 1+ u ).
\ee
The normalization ${\cal N}(u)$ is fixed by requiring
\be
Z_{BB'}^{(2)}(X_0-X_0')
= \langle B(u'); X_0' | e^{-lH} | B(u); X_0 \rangle.
\ee

The Neumann boundary state is
\be
| N \rangle = \lim_{u \rightarrow 0} | B(u) ; X_0 \rangle
= \left( 2\alpha' \right)^{-1/4}
\exp\left( - \sum_{n=1}^{\infty} \frac{1}{n}
\tilde{\alpha}_{-n} \alpha_{-n} \right) | 0 \rangle.
\ee
The Dirichlet boundary state is
\bel{BtoD}
| D ; X_0 \rangle =
\lim_{u \rightarrow \infty} | B(u) ; X_0 \rangle
=  ( 2\pi^2 \alpha' )^{1/4}
\exp\left( - i X_0 p \right)
\exp \left( \sum_{n=1}^{\infty} \frac{1}{n}
\tilde{\alpha}_{-n} \alpha_{-n}
\right) | 0 \rangle \! \rangle.
\ee
Here
$| 0 \rangle \! \rangle = | X=0 \rangle \otimes | 0 \rangle_{\alpha}$.

It is easy to evaluate the $g$-function
\be
g(u) := \langle 0 | B(u) ; X_0 \rangle
= ( \alpha'/2 )^{1/4} (2\pi u)^{-1/2}
\Gamma(u+1) (e/u)^u.
\ee
In the IR limit,
$g_{IR} = g(\infty) = \langle 0 | D ; X_0 \rangle =
( \alpha'/2)^{1/4}$,
and in the UV limit,
$g_{UV} = g(0) = \langle 0 | N \rangle
= \left( 2\alpha' \right)^{-1/4} \delta(0) = \infty$.
We see
$g_{UV}/g_{IR} = \left( \alpha' \right)^{-1/2} \delta(0) = \infty$.
This is consistent with the $g$-theorem \cite{AL}.


\resection{SBQD model on a cylinder}

We propose that the action of the supersymmetric BQD
model on the cylinder is given by
\bel{ASBQD}
S = \frac{1}{2\pi} \int_0^{2\pi r} {\rm d}\sigma^2
\int_0^l {\rm d}\sigma^1 {\cal L}_0 +
\frac{1}{4\pi} \int_0^{2\pi r} {\rm d}\sigma^2 {\cal L}_b,
\ee
where
\ba
{\cal L}_0 &=& \frac{2}{\alpha'} \partial_{\bar{z}} X \partial_z X
+ \psi \partial_{\bar{z}} \psi
+ \tilde{\psi} \partial_z \tilde{\psi}, \cr
{\cal L}_b &=&  a_1 \partial_2 a_1
+ h \omega_0^{-1} a_1 \left( \psi + i \tilde{\psi} \right)
+ \frac{1}{2\alpha'} h^2 \left( X - X_0 \right)^2 + i \tilde{\psi} \psi
\Bigr|_{\sigma^1=0} \nn
& & +  a_2 \partial_2 a_2
+ h' \omega_0^{-1} a_2 \left( \psi - i \tilde{\psi} \right)
+ \frac{1}{2\alpha'} (h')^2 \left( X - X'_0 \right)^2 + i \tilde{\psi} 
\psi
\Bigr|_{\sigma^1=l}.
\ea
Here $\omega_0 = e^{i \pi /4}$. $h, h' \in {\bf R}$.

The action is a direct sum of bosonic part and fermionic part:
$S = S[X] + S[\psi, \tilde{\psi}]$. Thus the partition function of the
SBQD model factorizes as
$Z = Z_{BB'}^{(X)} Z^{(\psi)}(h, h')$.

 From this action, we have the following boundary conditions
\be
\partial_1 X - \frac{1}{2} h^2 \left( X - X_0 \right) 
\Bigr|_{\sigma^1=0} =
0,
\qq
\partial_1 X + \frac{1}{2} (h')^2 \left( X - X'_0 \right)
\Bigr|_{\sigma^1=l} = 0,
\ee
\be
\psi - i \tilde{\psi} + h \omega_0^{-1} a_1 \Bigr|_{\sigma^1=0} = 0,
\qq
\psi + i \tilde{\psi} + h' \omega_0^{-1} a_2 \Bigr|_{\sigma^1=l} = 0,
\ee
\be
  2 \partial_2 a_1
- h \omega_0^{-1} \left( \psi +i  \tilde{\psi} \right)
\Bigr|_{\sigma^1=0} = 0,
\qq
  2 \partial_2 a_2
+ h' \omega_0^{-1} \left( \psi - i \tilde{\psi} \right)
\Bigr|_{\sigma^1=l} = 0.
\ee
After integrating over the auxiliary fermionic fields
$a_1(\sigma^2)$ and $a_2(\sigma^2)$,
the boundary conditions become
\be
\partial_1 X - v \left(  X - X_0 \right) \Bigr|_{\sigma^1=0} = 0, \qq
\partial_1 X + v' \left(  X - X'_0 \right) \Bigr|_{\sigma^1=l} = 0,
\ee
\be
\left( \partial_2 - i v \right) \psi \Bigr|_{\sigma^1=0}
= i \left( \partial_2 + i v \right) \tilde{\psi} \Bigr|_{\sigma^1=0}, \qq
\left( \partial_2 + i v' \right) \psi \Bigr|_{\sigma^1=l}
= -i \left( \partial_2 - i v' \right) \tilde{\psi} \Bigr|_{\sigma^1=l}.
\ee
Here $v = (1/2) h^2$, $v' = (1/2) (h')^2$.
With this identification of parameters, the bosonic part is
 the same as that of the previous section.

 From now on, we consider the fermionic part only.
Let us expand
\be
\psi(z) = \sum_{l} b_l e^{ik_l z}, \qq
\tilde{\psi}(\bar{z}) = \sum_{l} \tilde{b}_l e^{-ik_l \bar{z}}.
\ee
Then
\be
\tilde{b}_l = - i \left( \frac{k_l + i v}{k_l - i v} \right) b_l
=  i \left( \frac{k_l - i v'}{k_l + i v'} \right) e^{2i k_l l} b_l.
\ee
Thus the momentum $k_l$ must satisfy
\be
e^{2i k_l l} \left( \frac{k_l - i v}{k_l + i v} \right)
\left( \frac{k_l - i v'}{k_l + i v'} \right) + 1 = 0.
\ee


\subsection{Spectral determining function}

Let $\{ \lambda_j \}$ be a set of positive solutions of
$F_{BB'}^{(+)}(k)=0$ where
\bel{fbbp}
F_{BB'}^{(+)}(k) = \frac{1}{2}
\left[
1 + e^{2\pi i k}
\left( \frac{\pi k - i w}{\pi k + i w} \right)
\left( \frac{\pi k - i w'}{\pi k + i w'} \right)
\right].
\ee
This function has two poles at
$\pi k = -iw$ and $\pi k = -iw'$ and infinite number of zeros
at $k = \pm \lambda_j$.
We can set the order of $\lambda_j$ as
\be
0 \leq \lambda_1 < \lambda_2 < \ldots < \lambda_j < \lambda_{j+1} < 
\ldots.
\ee

The function (\ref{fbbp}) can be rewritten as
\bal{FpBB}
F^{(+)}_{BB'}(k)
&=& \frac{e^{\pi i k}}{(\pi k + i w)(\pi k + i w')}
\left[
\left( \pi^2 k^2 - w w' \right) \cos(\pi k)
+ ( w + w' ) \pi k \sin (\pi k)
\right] \nn
&=& \frac{ e^{\pi i k}}
{(1- i (\pi k/w))(1- i (\pi k/w'))}
\prod_{j=1}^{\infty}
\left[ 1 - \frac{k^2}{\lambda_j^2} \right].
\ea

(i) For DD ($w \rightarrow \infty$ and $w' \rightarrow \infty$),
\be
F_{DD}^{(+)}(k)
= \frac{1}{2} \left( 1 + e^{2\pi i k} \right)
= e^{\pi i k} \cos \pi k
= e^{\pi i k} \prod_{j=1}^{\infty}
\left( 1 - \frac{k^2}{(j-1/2)^2} \right).
\ee
\be
\lambda_j = j - \frac{1}{2}, \qq j=1, 2, \ldots.
\ee

(ii) For DN or ND ($w \rightarrow \infty$ and $w' \rightarrow 0$,
or $w \rightarrow 0$ and $w' \rightarrow \infty$),
\be
F_{DN}^{(+)}(k) =F_{ND}^{(+)}(k)
= \frac{1}{2} \left( 1 - e^{2\pi i k} \right)
= -i e^{\pi i k} \sin \pi k
= -i e^{\pi i k} \pi k \prod_{j=2}^{\infty}
\left( 1 - \frac{k^2}{(j-1)^2} \right).
\ee
\be
\lambda_1 = 0, \qq
\lambda_j = j - 1, \qq j=2, 3, \ldots.
\ee

(iii) For NN ($w \rightarrow 0$ and $w' \rightarrow 0$),
\be
F_{NN}^{(+)}(k)
= \frac{1}{2} \left( 1 + e^{2\pi i k} \right)
= e^{\pi i k} \cos \pi k
= e^{\pi i k} \prod_{j=2}^{\infty}
\left( 1 - \frac{k^2}{(j-3/2)^2} \right).
\ee
\be
\lambda_j = j - \frac{3}{2}, \qq j= 2, 3, \ldots.
\ee

There is a subtlety in the NN limit. Note that
\be
\lim_{w, w' \rightarrow 0}  \lambda_1 = 0,
\ee
\be
0 = \lim_{w, w' \rightarrow 0} F_{BB'}^{(+)}(\lambda_1)
  \neq \lim_{w, w' \rightarrow 0} F_{BB'}^{(+)}(0) = F_{NN}^{(+)}(0) = 1.
\ee
Although $F_{NN}^{(+)}(0) \neq 0$, it is convenient to set
$\lambda_1 =0$ at the NN point.

The other limits are
\ba
F_{BD}^{(+)}(k)
&=& \frac{1}{2}
\left[ 1 - e^{2\pi i k} \left( \frac{\pi k - i w}{\pi k + i w} \right)
\right]
= \frac{e^{\pi i k}}{(1- i(\pi k/w))}
\left[ \cos(\pi k) - \frac{\pi k}{w} \sin (\pi k) \right] \nn
&=& \frac{e^{\pi i k}}{( 1 - i (\pi k/w))}
\prod_{j=1}^{\infty}
\left[
1 - \frac{k^2}{\lambda_j^2}
\right].
\ea
\bal{FpBN}
F_{BN}^{(+)}(k)
&=& \frac{1}{2}
\left[ 1 + e^{2\pi i k} \left( \frac{\pi k - i w}{\pi k + i w} \right)
\right]
= - i \frac{\pi k e^{\pi i k}}{w - i \pi k}
\left[ \cos(\pi k) + w \frac{\sin \pi k}{\pi k} \right] \nn
&=&
- i \frac{\pi k(1+w)e^{\pi i k}}{(w - i \pi k)}
\prod_{j=2}^{\infty}
\left[
1 - \frac{k^2}{\lambda_j^2}
\right].
\ea
Comparing eq.(\ref{FpBB}) with eq.(\ref{FpBN}), we see that
for small $w$ or $w'$, $\lambda_1$ approaches zero as
\be
\pi \lambda_1 \sim
\left( \frac{ww'}{(1+w)(1+w')} \right)^{1/2},
\ee
and one of the zero points of $F_{BB'}(k)$ is cancelled with
a pole at $k=-iw/\pi$ or at $k=-iw'/\pi$.


\subsection{Zeta function regularization}

Let us regularize the following divergent products
by zeta function regularization.
\be
W_{BB'}^{(P)}:= \prod_{m \in {\bf Z}}
\prod_{j=1}^{\infty}
\left[
\left( \frac{m}{r} \right)^2
+ \left( \frac{\pi \lambda_j}{l} \right)^2
\right]^{1/2},
\ee
\be
W_{BB'}^{(A)}:= \prod_{s \in {\bf Z}+1/2}
\prod_{j=1}^{\infty}
\left[
\left( \frac{s}{r} \right)^2
+ \left( \frac{\pi \lambda_j}{l} \right)^2
\right]^{1/2}.
\ee
Similarly, by using the property of $F_{BB'}^{(+)}(k)$, we can obtain the
following regularized expressions:
\ba
W_{BB'}^{(P)} &=& q^{-(1/2)c_{BB'}^{(+)}}
\prod_{j=1}^{\infty} \left( 1 - q^{\lambda_j} \right) \nn
&=& \frac{\sqrt{8\pi^2 u u'}}{\Gamma(u+1) \Gamma(u'+1)}
\left( \frac{u}{e} \right)^u
\left( \frac{u'}{e} \right)^{u'} \nn
& & \ \ \times
\tilde{q}^{1/24}
\prod_{m=1}^{\infty}
\left[
1 + \left( \frac{m-u}{m+u} \right)
\left( \frac{m-u'}{m+u'} \right) \tilde{q}^m
\right].
\ea
\ba
W_{BB'}^{(A)} &=& q^{-(1/2)c_{BB'}^{(+)}}
\prod_{j=1}^{\infty} \left( 1 + q^{\lambda_j} \right) \nn
&=&
\frac{2\pi}{\Gamma(u+1/2) \Gamma(u'+1/2)}
\left( \frac{u}{e} \right)^u
\left( \frac{u'}{e} \right)^{u'} \nn
& &  \ \ \times \tilde{q}^{-1/48}
\prod_{m=1}^{\infty}
\left[
1 + \left( \frac{m-(1/2)-u}{m-(1/2)+u} \right)
\left( \frac{m-(1/2)-u'}{m-(1/2)+u'} \right) \tilde{q}^{m-1/2}
\right],
\ea
where
\be
c_{BB'}^{(+)} := \int_0^{\infty} {\rm d}t \ 2t \
\rho(it)
\left[ 1 + e^{2\pi t}
\left( \frac{\pi t + w}{\pi t - w} \right)
\left( \frac{\pi t + w'}{\pi t - w'} \right)
\right]^{-1}.
\ee

Here we have regarded the following infinite product
\be
\prod_{m=1}^{\infty} \left( 1 + \frac{u}{m-(1/2)} \right)
= \frac{\Gamma(u+1) e^{- \gamma u}}{\Gamma(2u+1)}
\prod_{m=1}^{\infty} e^{u/(m-(1/2))}
\ee
as
\be
\prod_{m=1}^{\infty} \left( 1 + \frac{u}{m-(1/2)} \right)
= \frac{\sqrt{\pi}}{\Gamma(u+1/2)}
\left( \frac{u}{e} \right)^u,
\ee
by replacing the infinite sum by
\be
\sum_{m=1}^{\infty} \frac{u}{ m - (1/2)}
\rightarrow
\lim_{s \rightarrow 1}
\left(u^s \zeta(s, 1/2) - \frac{su}{s-1} \right)
= \gamma u + u \log u - u + 2u \log 2.
\ee


\subsection{Fermionic Boundary State}

The boundary state for the fermionic sector is defined by relations
\bal{BSC}
\left( \partial_2 - i v \right) \psi|_{\sigma^1=0} | B(u) \rangle
&=& i \left( \partial_2 + i v \right) \tilde{\psi} \Bigr|_{\sigma^1=0}
| B(u) \rangle, \nn
\langle B(u) | \left( \partial_2 + i v \right) \psi|_{\sigma^1=0}
&=& -i \langle B(u) |
\left( \partial_2 - i v \right) \tilde{\psi} \Bigr|_{\sigma^1=0}.
\ea

The mode expansion is given by
\be
\psi(z) = \sum_s \psi_s \frac{1}{\sqrt{r}} e^{-sz/r}, \qq
\tilde{\psi}(\bar{z}) = \sum_s \tilde{\psi}_s \frac{1}{\sqrt{r}}
e^{-s\bar{z}/r}.
\ee
$s \in {\bf Z} + 1/2$ for NS and $s \in {\bf Z}$ for R.
\be
\{ \psi_s, \psi_{s'} \} = \delta_{s+s',0}, \qq
\{ \tilde{\psi}_s, \tilde{\psi}_{s'} \} = \delta_{s+s',0}.
\ee
The Hamiltonian of this system can be expressed by
using the zero-mode generators of the Virasoro algebras:
\be
H = \frac{1}{r} \left( L_0 + \tilde{L}_0 - \frac{1}{24} \right).
\ee
For NS sector,
\be
L_0 = \sum_{s>0} s \psi_{-s} \psi_s, \qq
\tilde{L}_0 = \sum_{s>0} s \tilde{\psi}_{-s} \tilde{\psi}_s,
\qq s \in {\bf Z} +1/2,
\ee
and the unique NS ground state is normalized as
$\langle 0 | 0 \rangle = 1$.

The boundary state for NS sector is given by
\be
| B(u) \rangle^{NS} = g_+(u)
\exp\left[ \sum_{s>0} i
\left( \frac{s-u}{s+u}\right) \psi_{-s} \tilde{\psi}_{-s}
\right] | 0 \rangle,
\ee
\be
{}^{NS} \langle B(u) |
= g_+(u) \langle 0 |
\exp\left[ \sum_{s>0} i
\left( \frac{s-u}{s+u}\right) \psi_{s} \tilde{\psi}_{s}
\right].
\ee
If we require
\be
W_{BB'}^{(A)} = {}^{NS} \langle B(u') | e^{-lH} | B(u) \rangle^{NS},
\ee
 the normalization is fixed  and we find
\be
g_+(u) =
\frac{\sqrt{2\pi}}{\Gamma(u+1/2)}
\left( \frac{u}{e} \right)^u.
\ee

For R sector,
\be
L_0 = \sum_{m=1}^{\infty} m \psi_{-m} \psi_m + \frac{1}{16}, \qq
\tilde{L}_0 = \sum_{m=1}^{\infty} m
\tilde{\psi}_{-m} \tilde{\psi}_m + \frac{1}{16},
\ee
and the ground states form the two-dimensional representation
of Clifford algebra.

Our convention is
\be
\psi_0 | \sigma \rangle = \frac{1}{\sqrt{2}} | \mu \rangle, \qq
\psi_0 | \mu \rangle = \frac{1}{\sqrt{2}} | \sigma \rangle,
\ee
\be
\tilde{\psi}_0 | \sigma \rangle = \frac{i}{\sqrt{2}} | \mu \rangle, \qq
\tilde{\psi}_0 | \mu \rangle = -\frac{i}{\sqrt{2}} | \sigma \rangle.
\ee
\be
\langle \sigma | \sigma \rangle = \langle \mu | \mu \rangle = 1, \qq
\langle \sigma | \mu \rangle = 0.
\ee
The states $|\sigma \rangle$ and $|\mu \rangle$ correspond
to the order field and the disorder field respectively.

Then, it holds that
$\psi_0 | \sigma \rangle = -i \tilde{\psi}_0  | \sigma \rangle$,
$\langle \sigma | \psi_0 = i \langle \sigma | \tilde{\psi}_0$.
Therefore, $|\sigma \rangle$ is chosen to satisfy the condition
(\ref{BSC}).

The boundary state for the R sector is given by
\be
| B(u) \rangle^R = g_-(u)
\exp\left[ \sum_{m=1}^{\infty} i
\left( \frac{m-u}{m+u}\right) \psi_{-m} \tilde{\psi}_{-m}
\right] | \sigma \rangle,
\ee
\be
{}^R \langle B(u) |
= g_-(u) \langle \sigma |
\exp\left[ \sum_{m=1}^{\infty} i
\left( \frac{m-u}{m+u}\right) \psi_{m} \tilde{\psi}_{m}
\right].
\ee

By requiring
\be
W_{BB'}^{(P)} = {}^{R} \langle B(u') | e^{-lH} | B(u) \rangle^{R},
\ee
we have
\be
g_-(u) = \frac{2^{1/4} \sqrt{2\pi u}}{\Gamma(u+1)}.
\ee
Thus, we obtain the same expressions of boundary states
and $g_{\pm}(u)$ as \cite{C,LMSS}.

Here we summarize the relations:
\be
W_{BB'}^{(A)} = {\rm Tr} \left( e^{- 2\pi r H_{BB'} } \right)
= {}^{NS} \langle B(u') | e^{-lH} | B(u) \rangle^{NS},
\ee
\be
W_{BB'}^{(P)} =
{\rm Tr} \left( (-1)^F e^{- 2\pi r H_{BB'} } \right)
={}^{R} \langle B(u') | e^{-lH} | B(u) \rangle^{R},
\ee

In \cite{GZ}, a part of the perturbation terms is identified with the
boundary spin operators
\be
\sigma_B^{(0)}(\sigma^2) =  \omega_0^{-1} a_1 ( \psi + i \tilde{\psi} )
\Bigr|_{\sigma^1 =0}, \qq
\sigma_B^{(l)}(\sigma^2) = \omega_0^{-1} a_2 ( \psi - i \tilde{\psi} )
\Bigr|_{\sigma^1=l}.
\ee
They are nonlocal with respect to the fermionic fields
$\psi$, $\tilde{\psi}$ and yield square root branch points.
Thus, the insertion of these operators at the boundaries
forms a complete boundary state which is a superposition of the
$NS$ and $R$ boundary states:
\bel{Bmu}
| B; h \rangle := \frac{1}{\sqrt{2} }
\left( | B(u) \rangle^{NS} + {\rm sign}(h) | B(u) \rangle^{R} \right).
\ee
The fermionic part of the partition function is given by
\bal{Zpsi}
Z^{(\psi)}(h, h')
&=& \frac{1}{2} \left[ {\rm Tr} \left( e^{- 2\pi r H_{BB'} } \right)
+ {\rm sign}(h h')
{\rm Tr} \left( (-1)^F e^{- 2\pi r H_{BB'} } \right)
\right] \nn
&=& \langle B; h' | e^{-lH} | B; h \rangle.
\ea
Here
$B = B(u)$, $B'=B(u')$ and $u = rv = (1/2) r h^2$,
$u' = r v' = (1/2) r (h')^2$.
It is easy to see that even number insertions of $\sigma_B$ gives
nonzero contribution to the partition function.

It is known that the Ising model has three conformally invariant
boundary conditions:
free ($f$), fixed up ($+$) and fixed down ($-$) \cite{Ca}.

Indeed, the corresponding boundary states are
\be
| f \rangle = | B; h =0 \rangle,
\qq
| \pm \rangle = | B; h = \pm \infty \rangle.
\ee
Thus the boundary state (\ref{Bmu}) intermediates
the free boundary condition and the fixed boundary conditions.

If the boundary magnetic fields $h$, $h'$ increase
from $0$ to $\pm \infty$,
the open channel partition function (\ref{Zpsi}) flows from
\be
Z^{(\psi)}(0,0) = Z_{ff} =
q^{-1/48} \prod_{n=0}^{\infty}
\left( 1 + q^{n+1/2} \right)
\ee
to
\be
Z^{(\psi)}(\pm \infty, \pm \infty) = Z_{\pm, \pm}
= \frac{1}{2}
\left[ q^{-1/48}
\prod_{n=0}^{\infty} \left( 1 + q^{n+1/2} \right)
+ q^{-1/48}
\prod_{n=0}^{\infty} \left( 1 - q^{n+1/2} \right)
\right],
\ee
\be
Z^{(\psi)}(\pm \infty, \mp \infty) = Z_{\pm, \mp}
= \frac{1}{2}
\left[ q^{-1/48}
\prod_{n=0}^{\infty} \left( 1 + q^{n+1/2} \right)
- q^{-1/48}
\prod_{n=0}^{\infty} \left( 1 - q^{n+1/2} \right)
\right].
\ee

\resection{Discussion}

In this article, we have obtained the cylinder
partition functions of  a few two-dimensional
field theory models from the technique of zeta function
 regularization.  
 ( For computation on geometries other than disc and
 cylinder, see \cite{IN}.)  
A subtraction procedure (renormalization) is introduced  in order to reproduce the
correct expression at the conformal points. 
 From the expression of the partition functions and
with the help of the boundary states, the corresponding $g$-functions
are determined. These are main results of this paper.

These results for the partition functions
should be proved by using the spectral zeta function
for the Laplacian $- \partial^2$. 
Some works for the zeta funcion regularization related to the mixed (Robin) boundary condition
are found in \cite{BFSV,BGvNV}.


\vspace{1cm}

{\bf Acknowledgments}\\
The authors would like to thank  T. Suyama for helpful discussions. 
We are very grateful to B. Stefanski, jr. for comments and for
pointing us ref. \cite{BK}.


\appendix

\resection{Appendix: Casimir energies}

In this appendix, we rewrite the spectral zeta functions
\be
\zeta_{BB'}^{(+)}(s) = \sum_{j=1}^{\infty} \frac{1}{\lambda_j^s}, \qq
\zeta_{BB'}^{(-)}(s) = \sum_{j=1}^{\infty} \frac{1}{\nu_j^s},
\qq {\rm Re} \ s>1
\ee
in an integral of Hermite type in order to
examine their values at $s=-1$.

Naively, Casimir energies $c_{BB'}^{(\pm)}$ were expected
to be $\zeta_{BB'}^{(\pm)}(-1)$.

Recall that
\be
F_{BB'}^{(\pm)}(k) = \frac{1}{2}
\left[
1 \pm e^{2\pi i k}
\left( \frac{\pi k - i w}{\pi k + i w} \right)
\left( \frac{\pi k - i w'}{\pi k + i w'} \right)
\right].
\ee
Note that
\be
\frac{\partial}{\partial k}
F_{BB'}^{(\pm)}(k) =
2\pi i \rho(k) \left[
F_{BB'}^{(\pm)}(k) - \frac{1}{2} \right],
\ee
where
\be
\rho(k) = 1 + \frac{w}{\pi^2 k^2 + w^2 }
+ \frac{w'}{\pi^2 k^2 + (w')^2 }.
\ee
\be
\frac{1}{2F_{BB'}^{(\pm)}(k)}
+ \frac{1}{2F_{BB'}^{(\pm)}(-k)} = 1.
\ee
Collecting these relations, we have
\be
U(k):= \frac{\partial}{\partial k}
\log F_{BB'}^{(\pm)}(k)
= 2\pi i \rho(k) \left[
1 - \frac{1}{2 F_{BB'}^{(\pm)}(k) } \right]
= 2\pi i \rho(k) \frac{1}{2 F_{BB'}^{(\pm)}(-k)}.
\ee
Explicitly, it can be written as
\be
\frac{\partial}{\partial k} \log F_{BB'}^{(+)}(k)
= i \pi - \frac{\pi}{\pi k + i w}
- \frac{\pi}{\pi k + i w'}
+ \sum_{j=1}^{\infty}
\left( \frac{1}{k - \lambda_j} + \frac{1}{k + \lambda_j} \right),
\ee
\be
\frac{\partial}{\partial k} \log F_{BB'}^{(-)}(k)
= i \pi + \frac{1}{k} - \frac{\pi}{\pi k + i w}
- \frac{\pi}{\pi k + i w'}
+ \sum_{j=1}^{\infty}
\left( \frac{1}{k - \nu_j} + \frac{1}{k + \nu_j} \right).
\ee
Let $\mu_j=\lambda_j$ ($\nu_j$) for $+$ ($-$).
For a natural number $M$,
let us choose a number $N$ such that $\mu_M < N < \mu_{M+1}$.

For simplicity, we assume $w, w'>0$. Then $\mu_1 >0$ and
we can choose a real number $\delta$ such that $0<\delta<\mu_1$.

Let an union of segments of the real axis $I$ be
\be
I = I_0 \cup I_1 \cup \ldots \cup I_M,
\ee
where
\be
I_0 ={[} \delta, \mu_1 - \epsilon {]},
\qq
I_j = {[} \mu_j + \epsilon, \mu_{j+1} - \epsilon {]}
\ \ (j=1, \ldots, M-1), \qq I_M = {[} \mu_M + \epsilon, N {]}.
\ee
Here $\epsilon$ is an infinitesimally small positive number.

For an analytic function $W(k)$
bounded on the strip $ 0 \leq {\rm Re}\  k \leq N$, we have
\be
0 = \int_{C_1} {\rm d}k \ U(k) W(k).
\ee
The integration contour $C_1$ is shown in Figure 1.
We get
\ba
0 &=&
\int_I {\rm d}k \ U(k) W(k)
+ i \int_0^R {\rm d}t \ U(N+it) W(N+it)
- i \int_{\delta}^R {\rm d}t \ U(it) W(it) \nn
& & - i \int_0^{\pi/2} {\rm d} \theta \ \delta e^{i\theta}
U(\delta e^{i\theta}) W(\delta e^{i\theta})
- \int_0^N {\rm d}t \ U(iR + t) W(iR +t) \nn
& & - \sum_{j=1}^M i \int_0^{\pi} {\rm d} \theta \
\epsilon e^{i\theta} U(\mu_j + \epsilon e^{i \theta})
W(\mu_j + \epsilon e^{i \theta}).
\ea

If we take $R \rightarrow \infty$ limit, we have
\bal{zC1}
& & i \pi \sum_{j=1}^M W(\mu_j) + {\cal O}(\epsilon) \nn
&=& \int_I {\rm d}k \ U(k) W(k)
- i \int_{\delta}^{\infty} {\rm d}t\ U(it) W(it)
+ i \int_0^{\infty} {\rm d}t U(N+it) W(N+it) \nn
& & - i \int_0^{\pi/2} {\rm d}\theta \ \delta e^{i \theta}
U(\delta e^{i\theta}) W(\delta e^{i\theta}).
\ea
Similarly, from
\be
0 = \int_{C_2} {\rm d}k \ U(-k) W(k),
\ee
(see Figure 1 for the integration contour $C_2$), we have
\bal{zC2}
& & i \pi \sum_{j=1}^M W(\mu_j) + {\cal O}(\epsilon) \nn
&=& \int_I {\rm d}k \ U(-k) W(k)
+ i \int_{\delta}^{\infty} {\rm d}t\ U(it) W(-it)
- i \int_0^{\infty} {\rm d}t U(-N+it) W(N-it) \nn
& & - i \int_0^{\pi/2} {\rm d}\theta \ \delta e^{-i \theta}
U(-\delta e^{-i\theta}) W(\delta e^{-i\theta}).
\ea
Adding eq.(\ref{zC1}) and eq.(\ref{zC2}) gives
\ba
   \sum_{j=1}^M W(\mu_j)
&=&  \int_{\delta}^N {\rm d}k \ \rho(k) W(k)
- \frac{1}{2\pi} \int_{\delta}^{\infty} {\rm d}t\ U(it)
\left[ W(it) - W(-it) \right] \nn
& &
+ \frac{1}{2\pi} \int_0^{\infty} {\rm d}t
\left[ U(N+it) W(N+it) - U(-N+it) W(N-it) \right] \nn
& & - \frac{1}{2\pi} \delta \int_0^{\pi/2} {\rm d}\theta \
\left[
e^{i \theta}
U(\delta e^{i\theta}) W(\delta e^{i\theta})
+ e^{-i \theta}
U(-\delta e^{-i\theta}) W(\delta e^{-i\theta})\right].
\ea
Here we used a relation
\be
U(k) + U(-k) = 2\pi i \rho(k),
\ee
and took the $\epsilon \rightarrow 0$ limit.

Let us set
\be
W(k) = k^{-s} = | k|^{-s} e^{-s {\rm arg}(k)}, \qq
- \pi < {\rm arg}(k) < \pi.
\ee
Then for ${\rm Re}\ s>1$, we have (as $M \rightarrow \infty$)
\bal{zBB}
\zeta_{BB'}^{(\pm)}(s) &=&
\int_{\delta}^{\infty} {\rm d}k\ \rho(k) \frac{1}{k^s}
- 2 \sin\left( \frac{\pi}{2}s \right)
\int_{\delta}^{\infty} {\rm d}t \
\frac{t^{-s}}{\displaystyle 1 \pm
e^{2\pi t}
\left( \frac{\pi t + w}{\pi t - w} \right)
\left( \frac{\pi t + w'}{\pi t - w'} \right)}
\rho(it) \nn
& & - \frac{1}{2\pi} \delta^{1-s}
\int_0^{\pi/2}{\rm d}\theta
\left[
e^{-i(s-1)\theta} U(\delta e^{i\theta})
+ e^{i(s-1)\theta} U(- \delta e^{- i \theta})
\right].
\ea

The second and the third terms
in the right hand side of eq.(\ref{zBB})
make senses in the whole complex $s$-plane.
When $\delta \rightarrow 0$,
the third term vanishes
for ${\rm Re}\ s<0$.

Let us consider the first term:
\be
\int_{\delta}^{\infty} {\rm d}k \ \rho(k) k^{-s}
=\frac{1}{s-1} \delta^{-s+1} + V(\delta, w, s) + V(\delta, w', s).
\ee
Here
\be
V(\delta, w, s) = \int_{\delta}^{\infty} {\rm d}k\
\frac{1}{k^s} \frac{w}{\pi^2 k^2 + w^2}.
\ee
This integral makes sense for ${\rm Re}\ s>-1$.

Note that for ${\rm Re}\ s>-1$,
$V(\delta, 0, s) = 0$, $V(\delta, \infty, s) = 0$.

For $0 < w < \infty$, it is possible to change the
integration variable from $k$ to
$t = w^2/(\pi^2 k^2 + w^2)$,
and then $V(\delta, w, s)$ becomes
\ba
V(\delta, w, s) &=&
\frac{1}{2\pi}\left( \frac{\pi}{w} \right)^s
\int_0^{\eta} {\rm d}t\ t^{(s-1)/2} ( 1 - t )^{-(s+1)/2} \nn
&=& \frac{1}{\pi(s+1)}\left( \frac{\pi}{w} \right)^s
\eta^{(s+1)/2}
F\left( \frac{s+1}{2}, \frac{s+1}{2}; \frac{s+3}{2}; \eta \right),
\ea
where $
\eta = w^2/(\pi^2 \delta^2 + w^2)$
and $F(\alpha, \beta; \gamma; \eta)$ is the hypergeometric function.
With the help of properties of
the hypergeometric function, we can
see that $V(\delta, w, s)$ has simple poles at
\be
s = 1-2m, \qq m=1, 2, 3, \ldots.
\ee
For $-3< {\rm Re}\ s<1$, $\delta \rightarrow 0$ limit gives
\be
V(0, w,s)= \frac{1}{2 \sin((s+1)\pi/2)} \left( \frac{\pi}{w} \right)^s.
\ee
Thus we conclude that
$\zeta_{BB'}^{(\pm)}(-1)$ diverges for
$0 < w, w' < \infty$.

We drop these divergent terms by hand and assume that the
Casimir energies are given by
\be
c_{BB'}^{(\pm)} := \int_0^{\infty} {\rm d}t \ 2t \
\rho(it)
\left[ 1 \pm e^{2\pi t}
\left( \frac{\pi t + w}{\pi t - w} \right)
\left( \frac{\pi t + w'}{\pi t - w'} \right)
\right]^{-1}.
\ee
These are finite and give correct values at conformal points.
Indeed,
\be
\int_0^{\infty} {\rm d}t \ \frac{2t}{1 + e^{2\pi t} } = \frac{1}{24}, \qq
\int_0^{\infty} {\rm d}t \ \frac{2t}{1 - e^{2\pi t} } = -\frac{1}{12}.
\ee
Integrating by parts, we can see that the above expressions
of the Casimir energies are
consistent with other expressions \cite{C,APS}.

For SBQD model, $\zeta_{BB'}^{(\pm)}(-1)$
appears only in a factor
$q^{(\zeta_{BB'}^{(-)}(-1)-\zeta_{BB'}^{(+)}(-1))/2}$.
In this case, the divergent terms cancel with each other:
\be
\lim_{s \rightarrow -1}
\left(
\zeta_{BB'}^{(-)}(s) - \zeta_{BB'}^{(+)}(s)
\right)
= c_{BB'}^{(-)} - c_{BB'}^{(+)}.
\ee
So we don't need to discard the divergent terms.



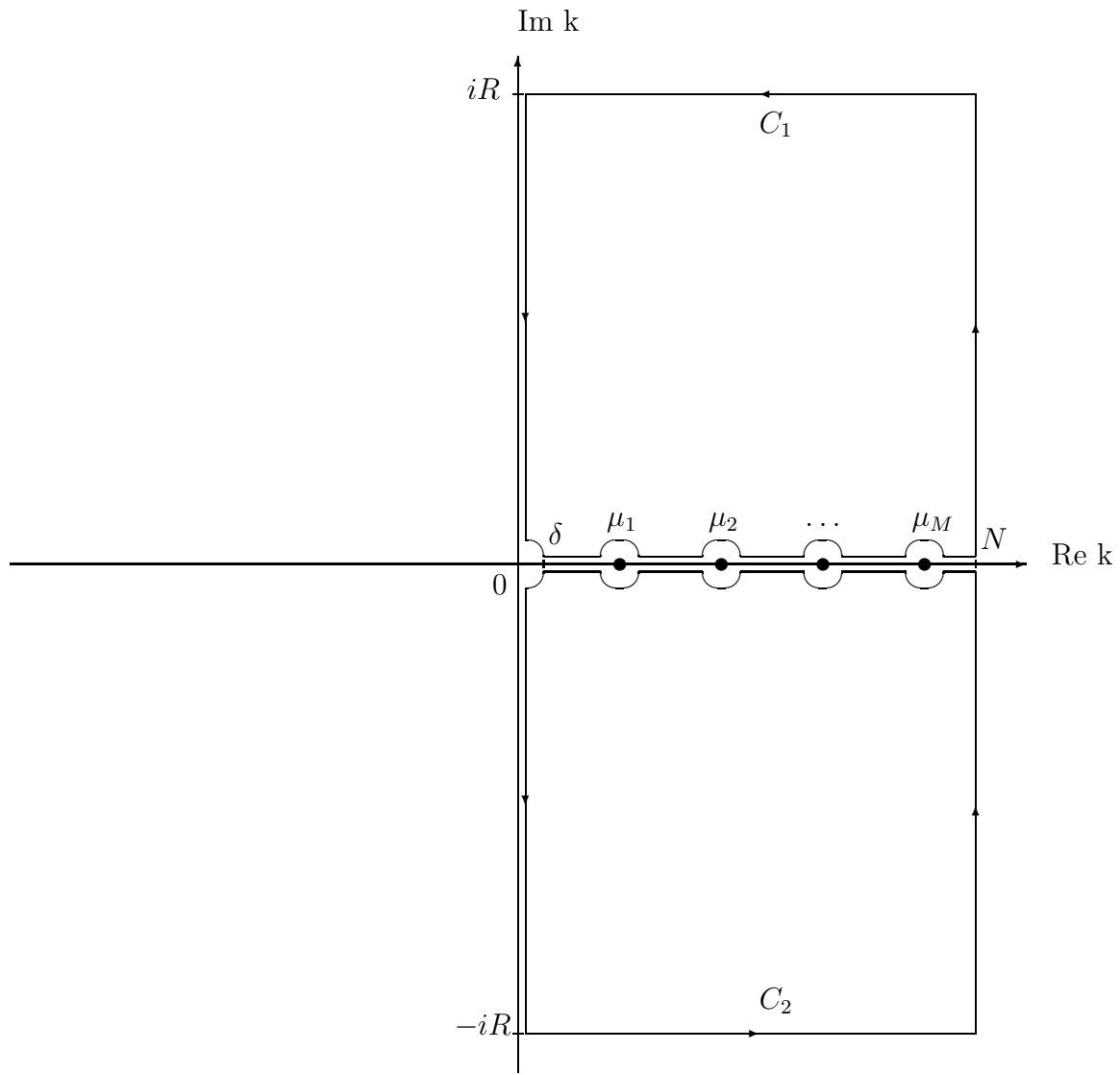
\begin{figure}[p]
\begin{picture}(400,400)
\put(0,200){\vector(1,0){400}}
\put(200,0){\vector(0,1){400}}
\put(410,200){{\rm Re} k}
\put(200,410){{\rm Im} k}
\put(190,188){$0$}
\put(210,198.5){\line(0,1){3}}
\put(212,208){$\delta$}
\put(380,198.5){\line(0,1){3}}
\put(382,206){$N$}
\put(198,15){\line(1,0){4}}
\put(175,15){$-i R$}
\put(198,385){\line(1,0){4}}
\put(180,383){$i R$}
\multiput(240,200)(40,0){4}{\circle*{5}}

\put(235,215){$\mu_1$}
\put(275,215){$\mu_2$}
\put(313,215){$\ldots$}
\put(355,215){$\mu_M$}
\put(295,370){$C_1$}
\multiput(240,203)(40,0){4}{\oval(15,13)[t]}
\put(203,203){\oval(14,13)[tr]}

\put(210,203){\line(1,0){22.5}}
\multiput(247.5,203)(40,0){3}{\line(1,0){25}}
\put(367.5,203){\line(1,0){12.5}}

\put(203,385){\vector(0,-1){90}}
\put(203,209.5){\line(0,1){85.5}}
\put(380,295){\line(0,1){90}}
\put(380,203){\vector(0,1){92}}
\put(203,385){\line(1,0){92}}
\put(380,385){\vector(-1,0){85}}
\put(295,25){$C_2$}
\multiput(240,197)(40,0){4}{\oval(15,13)[b]}
\put(203,197){\oval(14,13)[br]}

\put(210,197){\line(1,0){22.5}}
\multiput(247.5,197)(40,0){3}{\line(1,0){25}}
\put(367.5,197){\line(1,0){12.5}}

\put(203,190.5){\vector(0,-1){85.5}}
\put(380,197){\line(0,-1){92}}
\put(203,15){\line(0,1){90}}
\put(203,15){\vector(1,0){92}}
\put(295,15){\line(1,0){85}}
\put(380,15){\vector(0,1){90}}
\end{picture}
\caption{Contours of integration}
\end{figure}

\end{document}